# Beam splitters as controlled-$Z$ gate for hybrid state

Mikhail S. Podoshvedov[1,2,3] and Sergey A. Podoshvedov[1,2*]

[1]*Laboratory of quantum information processing and quantum computing, South Ural State University (SUSU), Lenin Av. 76, Chelyabinsk, Russia*
[2]*Laboratory of quantum engineering of light, South Ural State University (SUSU), Lenin Av. 76, Chelyabinsk, Russia*
[3]*Kazan Quantum Center, Kazan National Research Technical University named after A.N. Tupolev, Kazan, Russia*

[*] sapodo68@gmail.com

**Abstract:** We explore a scheme based on adding a nonlocal photon and subtracting some number of photons to entangle the initial single-mode squeezed vacuum (SMSV) state with the photon state. In a realistic model of interaction of the SMSV state with the photonic state on a beam splitter (BS) with changeable transmissivity or reflectivity the hybrid entanglement is realized for any values of the squeezing of input SMSV state. Maximum hybrid entanglement is achieved at certain values of the squeezing and BS parameter, which can mean implementation of a two-qubit $controlled-Z$ ($CZ-$) operation using the BS with the appropriate initialization of the input states. The success probability of the gate, taking into account multiphoton outcomes in the measuring mode of the BS, is more than 0.3. We also propose to use new continuous variable (CV) states of definite parity that could increase the success probability of generating maximal hybrid entanglement. We show sufficient robustness of the generated entanglement under photon number resolving detection with practical quantum efficiency.

**1. Introduction**

Idea of entangling two subsystems of different nature into unified system is often referred to as hybrid entanglement. A fairly common version of the optical hybrid entanglement involves both CV and discrete variable (DV) states, an idea that goes back to Schrödinger's pioneering work on the interpretation of quantum mechanics, so called Schrodinger's cat paradox [1]. The hybrid entanglement has been widely demonstrated in several optical systems [2-4] with an application for verification of the loophole-free violation of Bell's inequality [5]. In addition to its fundamental importance, the hybrid entanglement has a fairly wide range of practical applications. The hybrid entangled state can be used as a quantum channel in various modifications for transmission of unknown qubit [6-8]. It may become efficient resource for quantum key distribution [9] and quantum computation [10,11]. The hybrid entanglement can be used as a basis for building quantum networks [12,13] and also serve as converter [14]. The hybrid entangled states can also form the basis for conditional quantum engineering of new nonclassical states of light [15-19]. The method is based on measuring entangled system in auxiliary mode to control post-measurement state.

The hybrid entangled states are usually considered as a form of quantum correlations unifying coherent states of light with directly opposite amplitudes and a single photon, i.e. $|\alpha\rangle|0\rangle + |-\alpha\rangle|1\rangle$ [20], although researchers can deal with another type of the hybrid entangled states, obtained by replacing the coherent states with even and odd Schrödinger cat states (SCSs) $|SCS_\pm\rangle$, respectively, that is, $|SCS_-\rangle|0\rangle + |SCS_+\rangle|1\rangle$ [12,21]. Detecting a single photon in an indistinguishable manner, where the information that the photon is either subtracted from the SCS state, changing its parity, or simply detected, leaving the parity of the original state, is lost, allows the hybrid entanglement to be realized. Another type of the hybrid entanglement realized through superposition of the creation operators in neighboring modes, i.e. of the kind $a_1^+ + a_2^+$, is demonstrated in [22].

Photon subtracted and photon added CV states are experimentally realizable [23-27]. So, an application of the annihilation operator to the appropriate power $k$ to the Gaussian state $\rho$ allows for one to realize $k$ −photons subtracted state, i.e. $\rho_{-k} \sim a^k \rho a^{+k}$ [25]. But the annihilation and creation operators are non-unitary and cannot be directly implemented in the laboratory. Therefore, the action of the annihilation operator on the input state is approximated by passing the CV state through a high transmittance beam splitter with subsequent registration of $k$ photons in the adjacent measuring mode [25-27]. A realistic model of the interaction of the initial CV state with photonic states on the BS with variable transmissivity and reflectivity allows for one to accurately evaluate the output characteristics, in particular, to show the possibility of generating brighter nonclassical light [28]. Realistic approach provides the opportunity to take into account the redistribution of input photons for different BSs, involving highly transmitting one. So, the SMSV state is shown to be transformed into a hybrid entangled state if it passes through the BS [29]. Detection of a certain number of photons, given the development of photon resolving technology [30,31], conditionally generates the CV states of a certain parity, which are useful in quantum engineering of the SCS states [32,33].

Here we consider the combined action of adding a nonlocal photon to the SMSV state followed by subtracting some number of photons within a realistic interaction model to deterministically implement the hybrid entanglement. In the model, macroscopic objects are the CV states of a certain parity different from SMSV states, the average number of photons in which only increases with the increase in the number of subtracted photons [28]. The CV components are the states into which the initial SMSV state is transformed when passing through the BS with arbitrary transmission and reflection coefficients. Under certain conditions, the CV states can approximate SCSs states with high fidelity [32,33]. A realistic model allows us to estimate the probability of realizing the maximally hybrid entangled state of light in the case of multiphoton outcomes, with the exception of the vacuum outcome, the registration of which presents corresponding difficulties [34]. From a formal point of view, the generation of maximum hybrid entanglement can be interpreted as the action of the $CZ$ −gate implemented by BS. In realistic model, we estimate the fidelity of the output state in the case of using a photon number resolving (PNR) detector with non-unit quantum efficiency [30,31]. We also propose to use other input even CV states of a certain parity with smaller vacuum contribution as well as odd CV states to implement deterministic the hybrid entanglement.

## 2. Photon subtraction as a way to realize $CZ$ −operation with balanced photonic state

A scheme for the hybrid entanglement generation in Fig. 1 involves nonclassical resource in the form of the SMSV state [35] in mode 1

$$|SMSV(y)\rangle = \frac{1}{\sqrt{\cosh s}} \sum_{n=0}^{\infty} \frac{y^n}{\sqrt{(2n)!}} \frac{(2n)!}{n!} |2n\rangle, \tag{1}$$

and either a nonlocal single photon propagating in two separate spatial modes 2 and 3

$$|\xi\rangle_{23} = a_0 |01\rangle_{23} + a_1 |10\rangle_{23}, \tag{2}$$

or superposition of vacuum and correlated single photon states

$$|\varsigma\rangle_{23} = a_0 |00\rangle_{23} + a_1 |11\rangle_{23}. \tag{3}$$

For simplicity, we assume that the amplitudes of the states (2,3) $a_0$ and $a_1$ are real numbers satisfying the normalization condition $a_0^2 + a_1^2 = 1$. DV state in Eq. (2) can be realized by passing a single photon through the BS with transmittance and reflectance corresponding to the amplitudes of the nonlocal photon. The source of the photon state in Eq. (3) can be a two-mode squeezed vacuum (TMSV) state $|TMSV(s_1)\rangle = \sum_{n=0}^{\infty} tanh^n s_1 |nn\rangle/\cosh s_1$ [35] with small squeezing amplitude $s_1$. In the case of $s_1 = 0.1$, the amplitude of the correlated four-photon state $|22\rangle$ is approximately 100 times smaller than the amplitude of the state $|11\rangle$ which allows us to neglect higher-order correlated terms in TMSV states and limit oneself to consideration of the states as in Eq. (3). As for the SMSV state, a squeezing parameter $y = \tanh s/2$ is introduced to describe it with $s > 0$ being the squeezing amplitude which provides the range of its change $0 \leq y \leq 0.5$. In addition, the squeezing in decibels $S = -10 \log(exp(-2s))\, dB$ and the average number of photons $\langle n_{SMSV} \rangle = sinh^2 s$ can be used when describing the SMSV state. Practical squeezing values can range from $9\, dB$ [36] to $15\, dB$ [37]. In this section we consider the case of $a_0 = a_1 = 1/\sqrt{2}$.

The interaction of the SMSV state with second mode of the states in Eqs. (2,3) on the beam splitter the action of which is described by the matrix $BS_{12} = \begin{bmatrix} t & -r \\ r & t \end{bmatrix}$ with arbitrary real transmittance $t > 0$ and reflectance $r > 0$ subject to the normalization condition $t^2 + r^2 = 1$ with subsequent registration of Fock state in the measuring mode (mode 2) and replacing mode subscripts $(2 \leftrightarrow 3)$ generates, as shown in figure 1, the following hybrid entangled states

$$\left|\Delta_k^{(\xi)}(y_1, B)\right\rangle_{12} = \frac{1}{\sqrt{N_k}}\left(b_k|\Psi_k^{(1)}(y_1, B)\rangle_1 |0\rangle_2 + |\Psi_k^{(0)}(y_1, B)\rangle_1 |1\rangle_2\right), \quad (4)$$

$$\left|\Delta_k^{(\varsigma)}(y_1, B)\right\rangle_{12} = \frac{1}{\sqrt{N_k}}\left(|\Psi_k^{(0)}(y_1, B)\rangle_1 |0\rangle_2 + b_k|\Psi_k^{(1)}(y_1, B)\rangle_1 |1\rangle_2\right), \quad (5)$$

where the measurement-induced CV states of definite parity $|\Psi_k^{(0)}(y_1, B)\rangle$, $|\Psi_k^{(1)}(y_1, B)\rangle$ and related to them parameters: reduced squeezing parameter $y_1$, beam splitter parameter $B$, functions $Z(y_1)$, $Z^{(k)}(y_1)$ and $G_k^{(1)}(y_1, B)$ forming normalization factors of the corresponding CV states as well as the amplitudes $c_k^{(0)}(y_1, B)$ and $c_k^{(1)}(y_1, B)$ of the hybrid entangled state are presented in Appendix A. Here, the normalization coefficient of generated entangled state in Eqs. (4,5) is $N_k = 1 + b_k^2$ with an additional coefficient

$$b_k(y_1, B) = \frac{1}{\sqrt{1+B}} \begin{cases} \sqrt{B \frac{G_0^{(1)}(y_1)}{Z(y_1)}}, & if\ k = 0 \\ \frac{k}{\sqrt{y_1 B}}\sqrt{\frac{G_k^{(1)}(y_1, B)}{Z^{(k)}(y_1)}}, & if\ k > 0 \end{cases}, \quad (6)$$

arising as a result of the interaction of the SMSV state with the photonic state on the BS. Presented forms in equations (4,5) are provided by application of the phase shift operator on $\pi$, i.e. $P(\pi) = exp(i\pi a^+ a)$ in second mode in the case of $k \neq 0$. Probability distribution describing the conditional generation of the hybrid entangled states is the following

$$P_k(y_1, B) = \frac{c_k^{(0)2}(y_1, B)Z^{(k)}(y_1)N_k}{2\cosh s} = \frac{\sqrt{1-4y_1^2(1+B)^2}(y_1 B)^k Z^{(k)}(y_1) N_k}{2k!}. \quad (7)$$

Here, the orthogonality condition of the CV states $\langle \Psi_k^{(0)}(y_1, B)|\Psi_k^{(1)}(y_1, B)\rangle = 0$ is satisfied since the CV states are of opposite parity. Therefore, regardless of the values of the experimental input parameters $S$ and $B$, the generated entangled states can be considered in a four-dimensional Hilbert space. The negativity $\mathcal{N}$ has all required properties for the entanglement measure [38,39] and can be can be calculated in four-dimensional Hilbert space, that is, $\mathcal{N}_k = 2b_k/N_k$, where the case of $\mathcal{N}_k = 1$ arising in the case of $b_k = 1$ corresponds to the maximally entangled state. For the reason, the coefficient $b_k$ can also be called amplitude distorting as it can reduce the degree of entanglement of the generated state. The amplitude-distortion factor $b_k(y_1, B)$ depends on the values of the initial squeezing $S$ and the beam splitter parameter $B$, as $y_1$ can be expressed through $S$ and $B$. Since $b_k$ is not equal to zero for any values of $S$ and $B$, then, regardless of the measurement outcome in the measurement mode, the output state is always hybrid entangled, that is, one can talk about deterministic implementation of the hybrid entangled state.

Now we are going to use parity encoding to present the entangled states in equations (4,5) as output of a two-qubit gate. Indeed, the heralded CV states can be represented as even either $|+_{CV}\rangle = |\Psi_{2m}^{(0)}(y_1, B)\rangle$ for $k = 2m$ or $|+_{CV}\rangle = |\Psi_{2m+1}^{(1)}(y_1, B)\rangle$ for $k = 2m + 1$ and odd either $|-_{CV}\rangle = |\Psi_{2m+1}^{(0)}(y_1, B)\rangle$ for $k = 2m + 1$ or $|-_{CV}\rangle = |\Psi_{2m}^{(1)}(y_1, B)\rangle$ for $k = 2m$. As the basic elements of the computer zero and one, the corresponding superpositions of the $|+_{CV}\rangle$ and $|-_{CV}\rangle$ states can be used, i.e. CV states of indefinite parity, for example, $|0_{CV}\rangle = \left(|\Psi_{2m}^{(0)}(y_1, B)\rangle + |\Psi_{2m}^{(1)}(y_1, B)\rangle\right)/\sqrt{2}$ and $|1_{CV}\rangle = \left(|\Psi_{2m}^{(0)}(y_1, B)\rangle - |\Psi_{2m}^{(1)}(y_1, B)\rangle\right)/\sqrt{2}$ which are not explicitly applied here. We introduce a subscript $CV$ to distinguish them from photon superpositions. On the contrary, photonic states can be represented as superpositions of vacuum and a single photon, i.e. $|+\rangle = (|0\rangle + |1\rangle)/\sqrt{2}$ and $|-\rangle = (|0\rangle - |1\rangle)/\sqrt{2}$. Then, output entangled states in equations (4,5) can be formally realized by applying one-qubit rotation operator about the $y$ axis at the corresponding angle $\Theta_k$, that is, $R_y(\Theta_k) = \begin{bmatrix} cos(\Theta_k/2) & -sin(\Theta_k/2) \\ sin(\Theta_k/2) & cos(\Theta_k/2) \end{bmatrix} = \begin{bmatrix} t_k & -r_k \\ r_k & t_k \end{bmatrix}$, where the parameters of the qubit transformation are the following $t_k = (b_k + 1)/\sqrt{2N_k}$, $r_k = (b_k - 1)/\sqrt{2N_k}$ and functions $cos(\Theta_k/2)$ and $sin(\Theta_k/2)$ can be expressed through the coefficients $t_k$ and $r_k$, and $CZ -$ gate, where $Z$ is the Pauli matrix [40], i.e.

$$CZ R_{y,2}^T(\Theta_k) Z_1^{k+1}(|+_{CV}\rangle_1 |+\rangle_2), \qquad (8)$$

for the entangled state in equation (4) and

$$CZ R_{y,2}(\Theta_k) Z_1^k(|+_{CV}\rangle_1 |+\rangle_2), \qquad (9)$$

for the entangled state in equation (5), where the superscript $T$ indicates the transposition of the original matrix. Here an additional operation $Z$ to the power of either $k + 1$ or $k$ is used in order to allow for initialization $|+_{CV}\rangle_1 |+\rangle_2$. Thus, the measurement-induced hybrid entangled state can result of the $CZ -$gate action with a preliminary transformation of the initial photonic state associated with rotation around $y$ axis. For example, for $k = 2m$ in Eq. (8), we have the chain of transformations:
$CZ R_{y,2}^T(\Theta_k) Z_1^{2m+1}(|+_{CV}\rangle_1 |+\rangle_2) = Z_1 CZ R_{y,2}^T(\Theta_k)(|+_{CV}\rangle_1 |+\rangle_2) =$
$(b_{2m}|-_{CV}\rangle_1 |0\rangle_2 + |+_{CV}\rangle_1 |1\rangle_2)/\sqrt{N_{2m}}$,
which corresponds to the output state in equation (4) for $k = 2m$.

To get rid of the additional action of the rotation operator like $R_{y,2}(\Theta_k), R_{y,2}^T(\Theta_k)$ or the same from the amplitude distorting multiplier $b_k$ and realize alone $CZ -$gate, it is necessary to find the values of the parameters $S$ and $B$ that ensure the fulfillment of the condition $b_k = 1$ what guarantees $R_{y2}(\Theta_k) = I$, where $I$ is the unit operator. Then, the generation of the hybrid entangled states in equations (4,5) can be represented in the form of the output of the $CZ$ operation: either $CZ Z_1^{k+1}(|+_{CV}\rangle_1 |+\rangle_2)$ or $CZ Z_1^k(|+_{CV}\rangle_1 |+\rangle_2)$. Numerical findings show that the condition $b_k = 1$ can be satisfied only with even photon subtraction, i.e. for $k = 2m$, while for odd photon subtraction the parameters $S, B$ that ensure the fulfillment of the condition $b_{2m+1} = 1$ are not found for any $m$. At that the lines of unit values $b_{2m} = 1$ are horizontal in coordinates $S, B$, that is, with some $B = B_{2m,opt} = const$ the condition $b_{2m} = 1$ is satisfied at a given value of $S$, as shown in Figure 2(a-d) for $k = 0,2,4,6$, respectively. The horizontal lines represent level lines corresponding to a certain value of $B_{2m,opt}$. The maximum number of level lines is observed for $k=0$ in the lower part of the graph in Fig. 2(a), where the lines are located quite densely with some $B_{0,opt,max}$ being the maximum. Increasing the number of subtracted photons reduces the number of level lines to three for $k=4$ in Fig. 2(c) (say, $B_{4,opt,1}$, $B_{4,opt,2}$ and $B_{4,opt,3}$, where $B_{4,opt,3} > B_{4,opt,2} > B_{4,opt,1}$ and the third subscript is simply the serial number of the line), two in the case of $k=2$ in Fig. 2(b), i.e., $B_{2,opt,1}$ and $B_{2,opt,2}$, where $B_{2,opt,2} > B_{2,opt,1}$ and one $B_{opt,6}$ in Fig. 2(d) in the selected range of change $B$ leaving other possible level lines with larger $B$ not shown. It is interesting to note that $B_{0,opt,max} = B_{2,opt,2} = B_{4,opt,3} = B_{6,opt} = 8380.3$.

The main feature of the level lines is that they take values greater than 1, i.e. $B_{2m,opt} > 1$, which corresponds to the BS with predominance of reflectance $R = r^2$ over transmittance $T = t^2$, that is $R > T$. In the case of $B_{2m,opt} > 1000$, we can talk about use of a highly reflective beam splitter (HRBS) with $R \gg T$. Application of the HRBS in quantum engineering of new states is counterintuitive, quite often a highly transmitting beam splitter (HTBS) is used [15-17]. Use of the HRBS allows for one to redirect a significant part of the light energy into the measuring channel and thereby increase the success probability of the measurement outcomes. Values of the squeezing $S$ and $B_{k,opt} = B_{0,opt,max} = 8380.3$ in the used range of their changes and corresponding them the probabilities $P_k$ of the measurement outcomes following from equation (7) are presented in the Table 1.

The fact that the success probability $P_0$ is almost equal to 1, i.e., $P_0 \approx 1$, does not come as a surprise. Indeed, in the case of $s=0$, the input state in the first mode is a vacuum, and its mixing with a nonlocal single photon in equation (2) followed by measurement of the vacuum outcome, that is, the absence of a click, in the measuring mode nearly deterministically guarantees the heralded return of the photonic state in the case of $r \to 1$ and $t \to 0$. In addition, the amplitude $y_1$ of the output measurement-induced states $|\Psi_0^{(0)}(y_1, B)\rangle$ and $|\Psi_0^{(1)}(y_1, B)\rangle$ also approaches zero for $t \to 0$ transforming them into a near vacuum state. Taking into account it and also the fact that the practical registration of a no-click outcome is accompanied by some difficulties [34], we do not take vacuum contribution into account in the overall probability of outcomes that can generate maximal entangled hybrid state. As can be seen from the Table 1, equality $b_2 = b_4 = b_6 = 1$ can be performed under the same values of the parameters $S$ and $B_{2,opt,2} = B_{4,opt,3} = B_{6,opt}$, so, the total success probability of the events can be estimated as $P_{2,4,6} = P_2 + P_4 + P_6 = 0.342859$. Taking into account the measurement results of more than 6 photons, i.e. 8,10 and so on, and also by increasing the range of change of $S > 10\ dB$ it is possible to increase the total success probability to $\approx 0.36$. Accordingly, one can take the value for the probability of multiphoton implementation of the $CZ-$ gate by ideal PNR detector, i.e. $P_{CZ} = P_{2,4,6} \approx 0.36$. In general, the gate can be implemented with a significantly lower value of $B$, which entails some decrease in the probability $P_{CZ}$ of the $CZ$ gate implementation.

|  | $k=0$ | $k=2$ | $k=4$ | $k=6$ |
|---|---|---|---|---|
| $S$ | 0.0002 | 9.99995 | 9.99995 | 9.99995 |
| $B_{k,opt}$ | 8380.3 | 8380.3 | 8380.3 | 8380.3 |
| $P_k$ | 0.99994 | 0.19244 | 0.096619 | 0.05386 |

**Table 1.** Values $S$ and $B_{k,opt}$ which ensure not only the fulfillment of the condition $b_k = 1$ for $k = 0,2,4,6$ but also the highest possible success probabilities $P_k$ in the selected ranges of change $S$ and $B$. The values of $S$ for calculating $P_0$ and the remaining $P_2, P_4$ and $P_6$ lie at opposite ends of the range of change $S$.

## 3. Realization of controlled−Z gate with unbalanced photonic state

As noted in the previous section, implementation of the $CZ$ gate on click of PNR detector with input balanced DV state does not allow its realization, for example, with single photon outcome. To expand the possibilities for its implementation involving the cases with subtraction of odd photons, one should consider unbalanced photonic states in equations (2,3), use of which leads to the generation of the same hybrid entangled states as in equations (4,5) but with amplitude distorting coefficient $b'_k = a_1 b_k / a_1$ and the normalization factor $N'_k = 1 + b'^2_k$. The success probability of implementation of the $CZ-$ gate with preliminary corresponding transformation of the DV state, i.e. with $b'_k \neq 1$, is given by the formula (7) with replacement of the multiplier $N_k/2$ by $a_0^2 N'_k$ which is converted into $2a_0^2 = 2b_k^2/(1+b_k^2)$ in the case of perfect realization of two-qubit gate without prior single-qubit conversion, occurring when $b'_k = 1$ resulting in $N'_k = 2$, i.e.

$$P_k(y_1, B) = \sqrt{1 - 4y_1^2(1+B)^2} \frac{2(y_1 B)^k Z^{(k)}(y_1) b_k^2}{(1+b_k^2)k!}. \qquad (10)$$

Since there is an additional degree of control over the output hybrid entangled state, namely, relationship between $a_0$ and $a_1$, either $a_0/a_1$ or $a_1/a_0$, it enables to increase the success probability by amplifying multiplier $2b_k^2/(1+b_k^2)$ at least in the case of $b_k > 1$ on compared with one in equation (7). The amplitudes of the initial photonic states in Eqs. (2,3) are related to the corresponding parameter $b_k$ through the relations: $a_0 = b_k/\sqrt{1+b_k^2}$ and $a_1 = 1/\sqrt{1+b_k^2}$.

Figure 3(a) shows the dependences of even probabilities $P_0, P_2$ and $P_4$ on the initial squeezing $(S, dB)$ obtained by optimization according to the parameter $B$, i.e., for a given $S$, only those $B = B_{k,opt}$ are selected that ensures the maximum probability of the measurement outcome $P_{2m}$. Accordingly, figure 3(b) demonstrates the dependencies of $B_{0,opt}, B_{2,opt}, B_{4,opt}$ on $S$ that provide the success probabilities in the figure 3(a). As can be seen from the figure 3(b), all even optimizing values take the same values, i.e., $B_{0,opt} = B_{2,opt} = B_{4,opt} \approx 99$, which corresponds to use of the BS with greater reflectivity, i.e. with $R > T$. The values of even optimizing parameters $B_{2m,opt}$ differ from each other by thousandths, which leads to the appearance of a total horizontal line in Figure 3(b). It is interesting that the behavior of the optimizing values $B_{1,opt}, B_{3,opt}$ in figure 3(d) that provides the maximum probabilities of odd measurement outcomes $P_1, P_3$ in Figure 3(c) differs significantly from those presented in figure 3(b). For example, $B_{3,opt}$, although it takes on values greater than 1, can already reach the value $B_{3,opt} = 1$ with $S$ growing, which corresponds to use of the balanced beam splitter (BBS). On the contrary, the values of the optimizing parameter $B_{1,opt} < 1$ ($T > R$) with $B_{1,opt} \approx 1$ in the vicinity of $S \approx 0$ are observed. As for the probabilities, $P_0$ prevails over the others starting almost with $P_0 \approx 1$, but its contribution begins to fall with increasing $S$. The probabilities associated with the click of the PNR detector, i.e. $P_1, P_2, P_3$ and $P_4$ can increase with $S$ growing. The probability $P_1$ can especially significantly increase approaching $P_0$ from below with increase of $S$. The maximum observed probabilities $P_{k,max}$ for $k = 0,1,2,3,4$ and their accompanying values $S$ (although some of them with large squeezing $S$ may be difficult to implement in practice) and optimizing values $B_{k,opt}$, as well as the values of the parameter $b_k$ that allows us to find the amplitudes $a_0$ and $a_1$ of photonic state, are shown in Table 2.

|  | $k=0$ | $k=1$ | $k=2$ | $k=3$ | $k=4$ |
|---|---|---|---|---|---|
| $S, dB$ | 0.00017 | 24.535 | 9.981 | 29.996 | 12.464 |
| $B_{k,opt}$ | 99 | 0.01 | 99 | 0.01 | 99 |
| $P_{k,max}$ | 0.995 | 0.2616 | 0.184 | 0.081 | 0.0985 |
| $b_k$ | 0.995 | 1.709 | 1.852 | 1.669 | 1.75 |

**Table 2.** Maximum possible values of the success probabilities $P_{k.max}$ and values of the parameters $S$ and $B_{k,opt}$ in the selected range of their changes that provide them. Additionally, the values of the amplitude distorting factors $b_k$ for given values of $S$ and $B_{k,opt}$ which determine the amplitudes $a_0$ and $a_1$ of the DV state in Eqs. (2,3) are also demonstrated.

In order to estimate the probability of the multiphoton implementation of the $CZ-$gate excluding vacuum contribution in the case of an input unbalanced photonic state, it is necessary to find such $S$ and $B$ that could ensure the equality of the corresponding amplitude distorting factors $b_k$. If such values of the parameters exist, then the probability of the multiphoton $CZ-$gate can be estimated as the sum of the corresponding probabilities $P_k$. Our numerical results allow us to estimate the probability of multiphoton two-qubit transformation in three cases. Numerical findings in Figs. 3 show that achieving the maximum probability of even and odd outcomes requires different BSs with either $B>1$ for even or $B<1$ for odd outcomes. Therefore, the implementation of the $CZ-$ gate with both even and odd outcomes will most likely require use of BS with parameter $B$, a value of which is in the vicinity of 1. Indeed, in the case of $S=9.22\ dB$ and $B=0.46$ (despite the fact that the BS has more transmittance, it is already closer to balanced), we have $b_1=b_2=b_4=1$ which allows us to estimate the probability of the $CZ-$gate as $P_{CZ}=P_1+P_2+P_4=0.306$. Using BS with $B=0.359$ to mix the SMSV state with squeezing of $S=11,131\ dB$ with DV state allows us to implement $CZ-$gate with probability $P_{CZ}=P_1+P_3=0.284$, where $b_1=b_3=0.793$. Third case is associated with the values of $S=14,91\ dB$ and $B=0.135$, which allows the $CZ-$gate to be realized without a preliminary single-qubit transformation with probability $P_{CZ}=0.31$, where $b_1=b_3=0.5$. Other values of $S$ and $B$ allow the $CZ-$gate to be implemented only for one measurement event, the probabilities of which are presented in Fig. 3(a,c). Accordingly, all other measurement outcomes generate the hybrid entangled state with some amplitude distorting factor.

A factor that can have a rather destructive effect on increasing the probability of the multiphoton $CZ-$gate implementation is the significant contribution of the vacuum in the SMSV state. The amplitude of the vacuum state in the SMSV state is predominant in the case of original squeezing $<10\ dB$ which is the most common in practice [35]. As a consequence, the vacuum measurement outcome (no click) can dominate over the others in heralded generation of the hybrid entanglement. Therefore, it is quite in demand to propose a source of new CV states of a certain parity that would have smaller vacuum contribution. As such CV states, let us offer even CV state realized by subtracting two photons from the SMSV state that has previously passed through BS which follows from the formula (A2)

$$|\Psi_2^{(0)}(y_1,B)\rangle = \frac{1}{\sqrt{Z^{(2)}(y_1)}} \sum_{n=0}^{\infty} \frac{y_1^n}{\sqrt{(2n)!}} \frac{(2(n+1))!}{(n+1)!} |2n\rangle, \qquad (11)$$

and odd CV state that can be realized by adding a single photon to the SMSV one and subtracting two photons which stems from equation (A9)

$$|\Psi_2^{(1)}(y_1,B)\rangle = \sqrt{\frac{y_1}{G_2^{(1)}(y_1,B)}} \sum_{n=0}^{\infty} \frac{y_1^n}{\sqrt{(2n+1)!}} \frac{(2(n+1))!}{(n+1)!} \left(1 - \frac{2n+1}{2}B\right)|2n+1\rangle. \qquad (12)$$

The corresponding probability distributions $P_{SMSV}$, $P_{2,n}^{(0)}$ and $P_{2,n}^{(1)}$ of the CV states of both even SMSV and it from which two photons are subtracted in equation (11) and the odd CV state in expression (12) over number states $n$ are shown in Figure 4. As can be seen from the plots, the even two-photon subtracted CV state can have a significantly reduced vacuum contribution in $P_{2,n}^{(0)}$ due to an increase in the contributions from 2,4,6 and even 8 photons. In general, the probability distribution $P_{2,n}^{(0)}$ shifts towards multiphoton states and partly acquires a bell-shaped form in contrast to the original SMSV state. On the contrary, the odd probability distribution $P_{2,n}^{(1)}$ does not has a vacuum contribution and the maximum probability is fixed for single photon, that is for $P_{2,1}^{(1)}$. These circumstances can support their use in implementing the hybrid entanglement with higher probability of success.

## 4. Influence of quantum efficiency of PNR detector on quality of hybrid entanglement

The above estimates of the probability of the multiphoton implementation of the $CZ-$gate are carried out with PNR detector with unit fidelity. In realistic experiments, PNR detectors with a certain non-unit quantum efficiency $\eta \neq 1$ are used [33] which already leads to generation of a mixed state described by the density matrix presented in Appendix B, in particular, in Eq. (B2). Using the CV state generated by imperfect PNR detector, its fidelity and success probability to create it can be estimated as

$$Fid_k^{(\xi)} = \frac{1}{g_k^{(\xi)}}\left(1 + \frac{(1-\eta)^2}{2!}B^2\langle n_k^{(0)}\rangle\langle n_{k+1}^{(0)}\rangle\frac{N'_{k+2}}{N'_k}\left|\langle\Delta_{k+2}^{(\xi)}|\Delta_k^{(\xi)}\rangle\right|^2\right), \quad (13)$$

$$P_{k,\eta} = \eta^k g_k^{(\xi)} P_k, \quad (14)$$

where

$$g_k^{(\xi)} = 1 + (1-\eta)B\langle n_k^{(0)}\rangle\frac{N'_{k+1}}{N'_k} + \frac{(1-\eta)^2}{2}B^2\langle n_k^{(0)}\rangle\langle n_{k+1}^{(0)}\rangle\frac{N'_{k+2}}{N'_k} \quad (15)$$

and the success probability $P_k$ under use of perfect PNR detector is present in Eq. (10).

In Figure 5(a) we show the dependence of the fidelity of the hybrid entanglement in the case when even measurement outcomes are registered by PNR detector with $\eta = 0.6$, i.e., $Fid_2^{(\xi)}$ and $Fid_4^{(\xi)}$, on $S$. Here, the values of the parameters $S$ and $B$ are assumed to be chosen so that the $k-$photon-subtracted CV state $|\Delta_k^{(\xi)}\rangle$ is maximally entangled, i.e. with $b'_k = 1$ and $N'_k = 2$ but $b'_{k+j} \neq 1$ and $N'_{k+j} \neq 2$ for $j \neq 0$. As can be seen from the graphs, the more photons are subtracted, the faster the fidelity of the output hybridity decreases. The fidelity $Fid_2^{(\xi)}$ of the generated entanglement can be quite robust to fairly large errors in the detector determining the number of incoming photons. Accordingly, as can be seen from the figure 5(b), the success probability of generating the maximum hybrid entanglement decreases if it is implemented on the PNR detector with $\eta = 0.6$ compared to if it had been generated with ideal PNR detection $(P_2 > P_{2,\eta}, P_4 > P_{4,\eta})$. This is due to the increase in cases of not measuring $k$ incoming photons. In general, the probabilities $P_{2,\eta}, P_{4,\eta}$ can increase with increasing initial squeezing $S$ even if the PNR detector has non-ideal quantum efficiency $\eta = 0.6$.

## 5. Conclusion

The generation of the maximally entangled hybrid state [29] can find its practical application [6-19], therefore, in connection with advancement of photon-resolving measuring technologies [30,31], it is important to know the conditions under which it can be implemented more effectively taking into account the registration of the multiphoton outcomes [32,33] not limited to just a single-photon measurement outcome [25,26]. Here, we have investigated properties of generated hybrid entanglement and possible limits imposed on them when the SMSV state is subjected to the combined action of adding a nonlocal photon followed by subtracting a certain number of photons. The hybrid entanglement is constructed on the basis of the states of different parity for both CV states and photonic states, which allows for one to consider the generated entangled state in a four-dimensional Hilbert space. Due to the fact that the amplitude distorting factor does not take a zero value, the output entanglement is realized deterministically regardless of the values of the experimental parameters. From a formal point of view, the heralded implementation of the hybrid entanglement can be represented in terms of single-qubit operation performed on the photon state, which is associated with appearance of the amplitude distorting multiplier, with subsequent action of the two-qubit $CZ-$gate. The BS can act as $CZ-$gate [40] at certain values of the experimental parameters which ensure that the amplitude distortion factor is equal to one.

Note that the model with the BS with transmissivity close to unity can also show the possibility of realizing the hybrid entanglement. Indeed, replacing the beam splitter operator with the corresponding combination of creation and annihilation operators, one can obtain the following hybrid entangled state
$BS_{12}(|SMSV\rangle_1|\xi\rangle_{23}) \approx a_0 a_1^k a_2^{+k}(|SMSV\rangle_1|0\rangle_2)|1\rangle_3 + a_1 a_1^{k-1} a_2^{+k}(|SMSV\rangle_1|0\rangle_2)|0\rangle_3 = \left(a_0 a_1^k|SMSV\rangle_1|1\rangle_3 + a_1 a_1^{k-1}|SMSV\rangle_1|0\rangle_3\right)|k\rangle_2,$
when subtracting $k$ photons, where the CV states $a_1^k|SMSV\rangle$ and $a_1^{k-1}|SMSV\rangle$ are the states of different parity whose parity depends on the value of $k$. But such an approximation does not allow us to analyze the cases of using BBS, BSs with larger reflectivity and even more so HRBSs thus to estimate the probability of success of conditional generation of the hybrid entanglement. However, a realistic model of the interaction of the SMSV and nonlocal photonic states on the BS whose transmittance or reflectance can be changed allows us to accurately estimate the probability of successful implementation of the maximum hybrid entanglement. Taking into account the multiphoton outcomes that ensure the amplitude distortion factor being equal to one, the success probability of implementation of the $CZ-$gate excluding a vacuum contribution can be estimated higher of 0.3 for practically feasible values of the parameter $S$ [36,37] and $B$ that is, for the beam splitters that are not only either HTBSs or HRBSs. Taking into account the vacuum outcome can significantly increase the probability of implementing the $CZ-$gate, but from a practical point of view, setting up such a scheme can be quite complex [34]. Using the balanced photonic state allows to slightly improve the success probability by using only HRBSs. The hybrid entanglement generation can show some robustness in the case of detection with rather low quantum efficiency. The fidelity of the output entanglement decrease slightly with increasing initial squeezing.

In general, using even CV state of definite parity with smaller vacuum contribution compared to the SMSV state may become more efficient when implementing maximum hybrid entanglement with a higher probability of success. The odd CV state represented in equation (12) without vacuum contribution may also become useful when implementing alone $CZ-$gate since use of the states can significantly change the output redistribution of photons. Moreover, proposed CV states of definite parity in Eqs. (11,12) are brighter in comparison with the initial SMSV state. Application of the CV states deserves separate consideration.

**Appendix A. Interaction of SMSV state with photonic states (2,3)**

Here we present the main moments of interaction between the SMSV state in equation (1) with DV ones in Eqs. (2,3) on BS with arbitrary real transmittance $t > 0$ and reflectance $r > 0$. The theory is based on the simultaneous interaction of the CV state with the vacuum, that is, simply the passage of the SMSV state through the BS, and with a single photon. These cases are examined in turn.

The BS mixes the modes 1 and 2 transforming the creation operators $a_1^+$ and $a_2^+$ as $BS_{12}a_1^+BS_{12}^+ = ta_1^+ - ra_2^+$ and $BS_{12}a_2^+BS_{12}^+ = ra_1^+ + ta_2^+$, respectively. This transformation of the creation operators converts the original state into the hybrid entangled state [28,29,32,33]

$$BS_{12}(|SMSV(y)\rangle_1|0\rangle_2) = \frac{1}{\sqrt{\cosh s}}\sum_{k=0}^{\infty} c_k^{(0)}(y_1, B)\sqrt{Z^{(k)}(y_1)}|\Psi_k^{(0)}(y_1)\rangle_1 |k\rangle_2, \quad (A1)$$

with even ($k = 2m$)

$$|\Psi_{2m}^{(0)}(y_1)\rangle = \frac{1}{\sqrt{Z^{(2m)}(y_1)}}\sum_{n=0}^{\infty} \frac{y_1^n}{\sqrt{(2n)!}}\frac{(2(n+m))!}{(n+m)!}|2n\rangle \quad (A2)$$

and odd ($k = 2m+1$) CV components

$$|\Psi_{2m+1}^{(0)}(y_1)\rangle = \sqrt{\frac{y_1}{Z^{(2m+1)}(y_1)}}\sum_{n=0}^{\infty} \frac{y_1^n}{\sqrt{(2n+1)!}}\frac{(2(n+m+1))!}{(n+m+1)!}|2n+1\rangle \quad (A3)$$

whose amplitudes are given by

$$c_k^{(0)}(y_1, B) = (-1)^k \frac{(y_1 B)^{\frac{k}{2}}}{\sqrt{k!}}, \quad (A4)$$

where the input squeezing parameter $y$ decreases by $t^2$ times, that is, it becomes equal to $y_1 = yt^2 = y/(1+B) \leq y$ and the BS parameter is equal to $B = (1-t^2)/t^2$. The normalizing factors of the CV states of a certain parity in Eqs. (A2,A3) are determined through $2m, 2m+1$ derivatives of the analytical function $Z(y_1) = 1/\sqrt{1-4y_1^2}$, i.e., $Z^{(2m)}(y_1) = dZ^{2m}/dy_1^{2m}$ and $Z^{(2m+1)}(y_1) = dZ^{2m+1}/dy_1^{2m+1}$, respectively. The subscript $2m, 2m+1$ is responsible for the number of subtracted photons, while the superscript accounts for the number of additional input photons, so (0) corresponds to the input vacuum state. Note the average number of photons in the measurement-induced $k-$ photon subtracted CV states in Eqs. (A2,A3) are given by

$$\langle n_k^{(0)}\rangle = y_1 \frac{Z^{(k+1)}}{Z^{(k)}}. \quad (A5)$$

Adding single photon to original SMSV state can be realized by its presence at the entrance to the BS

$$BS_{12}(|SMSV(y)\rangle_1|1\rangle_2) = (ra_1^+ + ta_2^+)BS_{12}(|SMSV(y)\rangle_1|0\rangle_2), \quad (A6)$$

that generates the following hybrid entangled state

$$BS_{12}(|SMSV(y)\rangle_1|1\rangle_2) = \frac{1}{\sqrt{\cosh s}}\sum_{k=0}^{\infty} c_k^{(1)}(y_1, B)\sqrt{G_k^{(1)}(y_1, B)}|\Psi_k^{(1)}(y_1, B)\rangle_1 |k\rangle_2 \quad (A7)$$

with measurement-induced CV states of definite parity

$$|\Psi_0^{(1)}(y_1)\rangle = \frac{1}{\sqrt{G_0^{(1)}(y_1)}}\sum_{n=0}^{\infty} \frac{y_1^n}{\sqrt{(2n+1)!}}\frac{(2n)!}{(n)!}(2n+1)|2n+1\rangle, \quad (A8)$$

$$|\Psi_{2m}^{(1)}(y_1, B)\rangle = \sqrt{\frac{y_1}{G_{2m}^{(1)}(y_1, B)}}\sum_{n=0}^{\infty} \frac{y_1^n}{\sqrt{(2n+1)!}}\frac{(2(n+m))!}{(n+m)!}\left(1 - \frac{2n+1}{2m}B\right)|2n+1\rangle, \quad (A9)$$

$$|\Psi_{2m+1}^{(1)}(y_1, B)\rangle = \frac{1}{\sqrt{G_{2m+1}^{(1)}(y_1, B)}}\sum_{n=0}^{\infty} \frac{y_1^n}{\sqrt{(2n)!}}\frac{(2(n+m))!}{(n+m)!}\left(1 - \frac{2n}{2m+1}B\right)|2n\rangle, \quad (A10)$$

whose amplitudes become

$$c_k^{(1)}(y_1, B) = \frac{1}{\sqrt{1+B}}\begin{cases} \sqrt{B}, & \text{if } k = 0 \\ (-1)^{k+1}\frac{(y_1 B)^{\frac{k-1}{2}}}{\sqrt{k!}}k, & \text{if } k \neq 0 \end{cases}. \quad (A11)$$

The CV states have a more complex form in contrast to the measurement-induced CV ones of a certain parity without an additional input single photon, which leads to more extended expressions for the normalization factors

$$G_0^{(1)}(y_1) = \frac{d}{dy_1}(y_1 Z(y_1)) = Z^3(y_1), \tag{A12}$$

$$G_k^{(1)}(y_1, B) = Z^{(k-1)}(y_1) + a_{k,1}^{(1)}(B)\left(y_1 Z^{(k)}(y_1)\right) + a_{k,2}^{(1)}(B) y_1 \frac{d}{dy_1}\left(y_1 Z^{(k)}(y_1)\right), \tag{A13}$$

where

$$a_{k,1}^{(1)}(B) = -\frac{2B}{k}, \tag{A14}$$

$$a_{k,2}^{(1)}(B) = \left(\frac{B}{k}\right)^2. \tag{A15}$$

As in the case discussed above, the subscript $k$ indicates the number of photons being subtracted while the superscript $(1)$ is responsible for the number of added photons. Let us only note that the mean number of photons in the CV states in Eqs. (A8-A10) can be determined by a formula similar to formula (A5) but with the replacement of the normalization factors, that is,

$$\langle n_k^{(1)} \rangle = \frac{\left(y_1 \frac{d}{dy_1}\right) G_k^{(1)}(y_1, B)}{G_k^{(1)}(y_1, B)}. \tag{A16}$$

The approach with output hybrid entangled states is directly used when deriving target states in formulas (4,5). Let's consider them using balanced DV state in equation (2). Indeed, we have chain of transformations

$$BS_{12}(|SMSV(y)\rangle_1 |\xi\rangle_{23}) = \frac{1}{\sqrt{2}} \begin{pmatrix} BS_{12}(|SMSV(y)\rangle_1 |0\rangle_2)|1\rangle_3 + \\ BS_{12}(|SMSV(y)\rangle_1 |1\rangle_2)|0\rangle_3 \end{pmatrix}, \tag{A17}$$

which, using formulas (A1, A6), can finally be rewritten as

$$BS_{12}(|SMSV(y)\rangle_1 |\xi\rangle_{23}) = \frac{1}{\sqrt{2\cosh s}} \sum_{k=0}^{\infty} c_k^{(0)}(y_1, B) \sqrt{Z^{(k)}(y_1) N_k} \left|\Delta_k^{(\xi)}(y_1, B)\right\rangle_{13} |k\rangle_2. \tag{A18}$$

The measurement of the measuring mode (mode 2) generates the hybrid entangled state of type (4) with the probability given by formula (7). The state contains an additional coefficient

$$b_k(y_1, B) = \frac{c_k^{(1)}(y_1, B)}{c_k^{(0)}(y_1, B)} \sqrt{\frac{G_k^{(1)}(y_1, B)}{Z^{(k)}(y_1)}}, \tag{A19}$$

the analytical expression of which in Eq. (6) can be derived on the basis of expressions for the amplitudes $c_k^{(0)}(y_1, B)$ (A4) and $c_k^{(1)}(y_1, B)$. This coefficient indicates the preliminary use of a unitary transformation over the original photonic state associated with rotation around the axis $y$. A similar method can be used in the case of other photonic states including unbalanced ones in Eqs. (4,5).

**Appendix B. Derivation of the output state with imperfect PNR detector**

Effect of the quantum efficiency $\eta$ of PNR detector can be modeled by means of use of positive operator-valued measure (POVM) formalism $\{\Pi_k, k = 0, \ldots, \infty\}$ with POVM elements [33]

$$\Pi_k = \eta^k \sum_{x=0}^{\infty} \begin{pmatrix} C_{2(m+x)}^{2m}(1-\eta)^{2x} + |2(m+x)\rangle\langle 2(m+x)| + \\ C_{2(m+x)+1}^{2m}(1-\eta)^{2x+1} + |2(m+x)+1\rangle\langle 2(m+x)+1| \end{pmatrix}, \tag{B1}$$

where $C_{2(m+x)}^{2m}$ and $C_{2(m+x)+1}^{2m}$ are the binomial coefficients and $\eta$ is a quantum efficiency of the PNR detector. The case $\eta = 1$ corresponds to a perfect PNR detector.

Using the general definition of the measurement-induced state realized with corresponding measurement operator in Eq. (B1) and taking into account terms up to the second order of smallness on parameter $\eta$, i.e. $(1-\eta)^2$, for example for input photonic state in Eq. (2), we have

$$\rho_k^{(\xi)} = \frac{1}{g_k^{(\xi)}} \begin{pmatrix} |\Delta_k^{(\xi)}\rangle\langle\Delta_k^{(\xi)}| + (1-\eta)B\langle n_k^{(0)}\rangle \frac{N'_{k+1}}{N'_k} |\Delta_{k+1}^{(\xi)}\rangle\langle\Delta_{k+1}^{(\xi)}| + \\ \frac{(1-\eta)^2}{2} B^2 \langle n_k^{(0)}\rangle\langle n_{k+1}^{(0)}\rangle \frac{N'_{k+2}}{N'_k} |\Delta_{k+2}^{(\xi)}\rangle\langle\Delta_{k+2}^{(\xi)}| \end{pmatrix}, \quad (B2)$$

where the multiplier $g_k^{(\xi)}$ is given in Eq. (13) and all the notations used are presented above.

**Acknowledgments**

The work was supported by the Foundation for the Advancement of Theoretical Physics and Mathematics "BASIS".

34. S.U. Shringarpure and J.D. Franson, "Generating photon-added states without adding a photon," Phys. Rev. A **100**, 043802 (2019).
35. R. Schnabel, "Squeezed states of light and their applications in laser interferometers," Phys. Reports **684** 1-51 (2017).
36. Y. Takeno, M Yukawa, H, Yonezawa and A. Furusawa, "Observation of -9 dB quadrature squeezing with improvement of phase stability in homodyne measurement," Opt. Express, **15** 4321-4327 (2007).
37. H. Vahlbruch, M. Mehmet, K. Danzmann and R. Schnabel, "Detection of 15 dB squeezed states of light and their application for the absolute calibration of photoelectric quantum efficiency," Phys. Rev. Lett. **117**, 110801 (2016).
38. A. Peres, "Separability criterion for density matrices," Phys. Rev. Lett. **77**, 1413-1415 (1996).
39. G. Vidal and R.F. Werner, "Computable measure of entanglement," Phys. Rev. A **65**, 032314 (2002).
40. M.A. Nielsen, "Cluster-state quantum computation," Rep. on Math. Phys. **57**, 147-161 (2006).

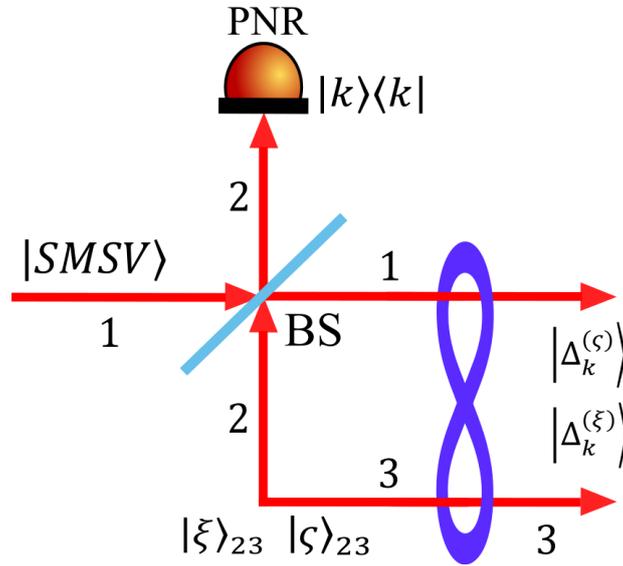

**Fig. 1.** Schematic representation of the optical scheme used to implement the hybrid entanglement based on the SMSV state and one of the two DV states in equation (2,3). The SMSV state is mixed with one of the modes of the two-mode photonic state, followed by measurement of the number of photons by the PNR detector, which deterministically generates the hybrid entanglement. Under certain conditions that is, for certain values of $S, B$ and amplitude relations of the DV states $a_1/a_0$, the BS used can implement $CZ-$gate with maximum hybrid entanglement at the output.

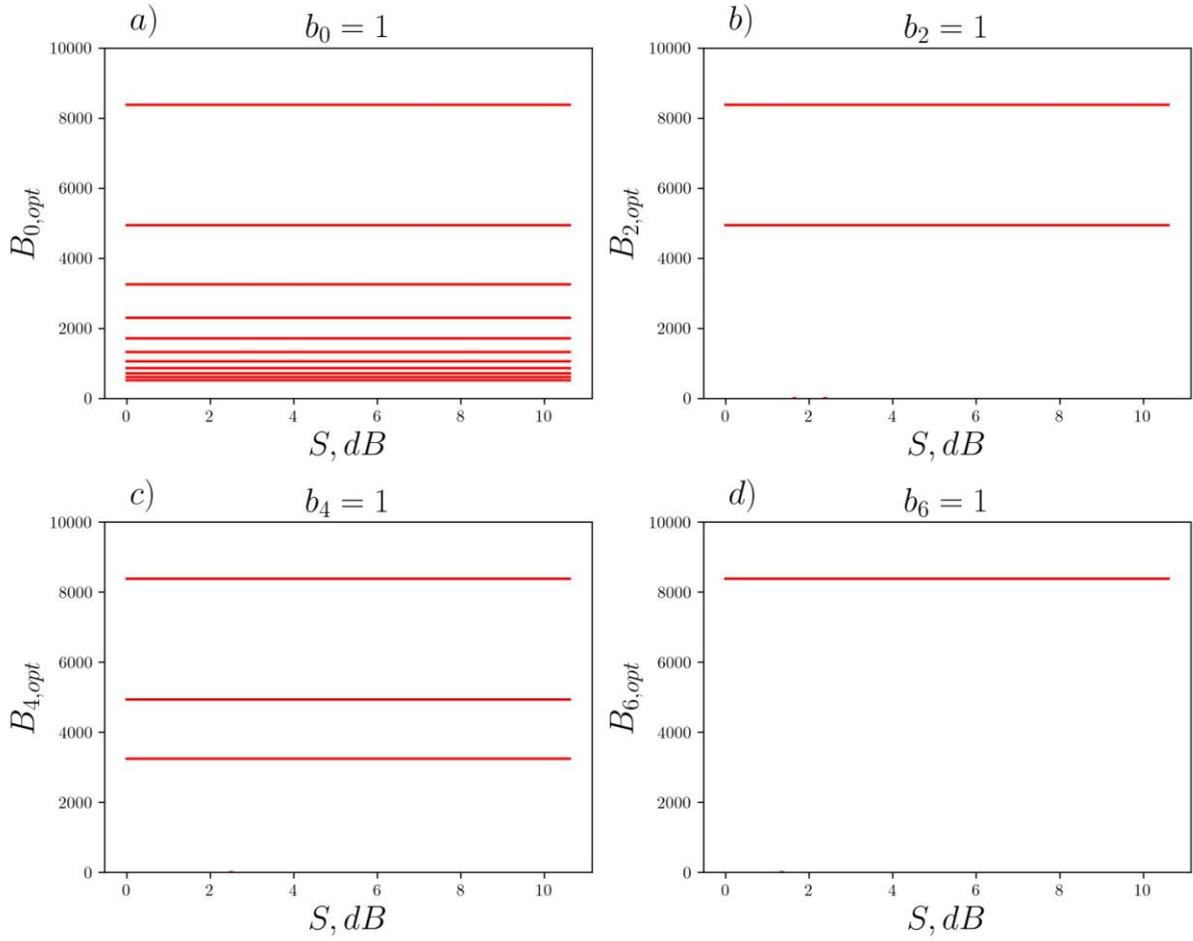

**Fig. 2(a-d).** Graphical dependencies of the BS parameter $B_{k,opt}$ and $S$ ($dB$) which ensure fulfillment of the condition $b_k = 1$ for the corresponding even numbers (a) $k = 0$, (b) $k = 2$, (c) $k = 4$ and (d) $k = 6$ in the case of use of the balanced DV state in equations (2,3). The dependencies are horizontal lines, and largest values of $B_{k,opt}$ can coincide for different values of $k$.

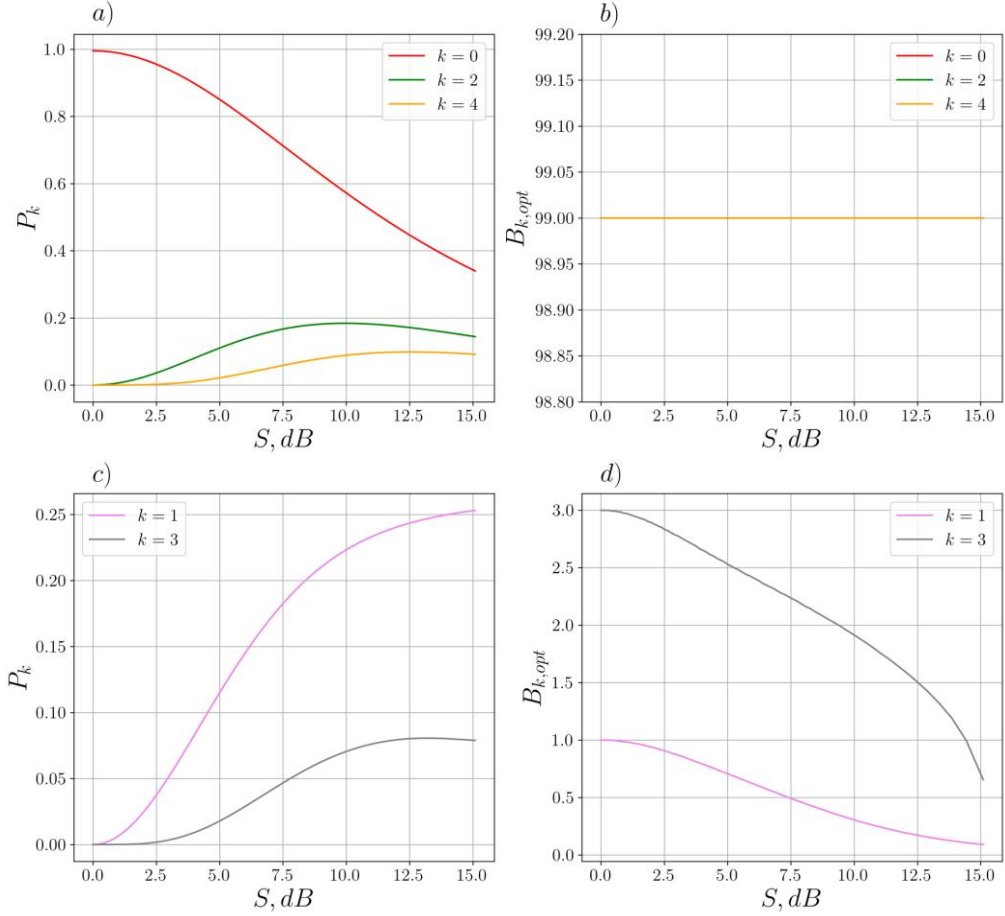

**Fig. 3(a-d).** (a) Dependences of the maximum even (a) probabilities $P_k$ for $k = 0,2,4$ and odd (c) ones for $k = 1,3$ of the measurement outcomes and, as a consequence, measurement-induced generation of the hybrid entangled states in equations (4,5) on the input squeezing $S$ of the SMSV state. The obtained probabilities are optimized by the BS parameter $B_{k,opt}$, dependencies of which on $S$ are shown for even (b) and odd (d) outcomes. The values of $B_{0,opt}, B_{2,opt}, B_{4,opt}$ almost coincide, differing from each other by thousandths, which is reflected on the graph in the form of a horizontal line.

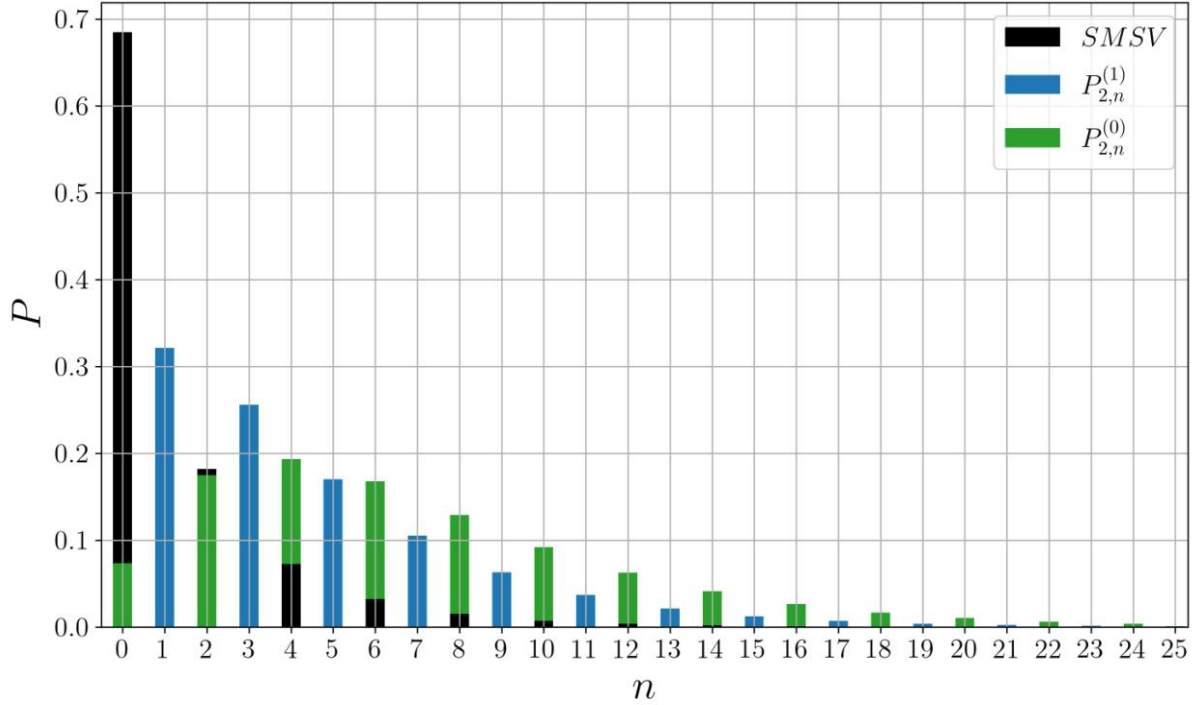

**Fig. 4.** Probability distributions for the CV states with two photons subtracted $P_{2,n}^{(0)}$ (even CV state) and a single photon added and two photons subtracted $P_{2,n}^{(1)}$. These CV states may be candidates for realizing the maximum hybrid entanglement with higher probability. For comparison, $SMSV$ distribution $P_{SMSV}$ with the maximum vacuum contribution is also presented. The distributions $P_{2,n}^{(0)}$ and $P_{2,n}^{(1)}$ are obtained with $S = 8\ dB$ taking into account optimization by the parameter $B$. The selected optimization provides maximum squeezing of noise of one of the quadrature components for the states in Eqs. (11,12).

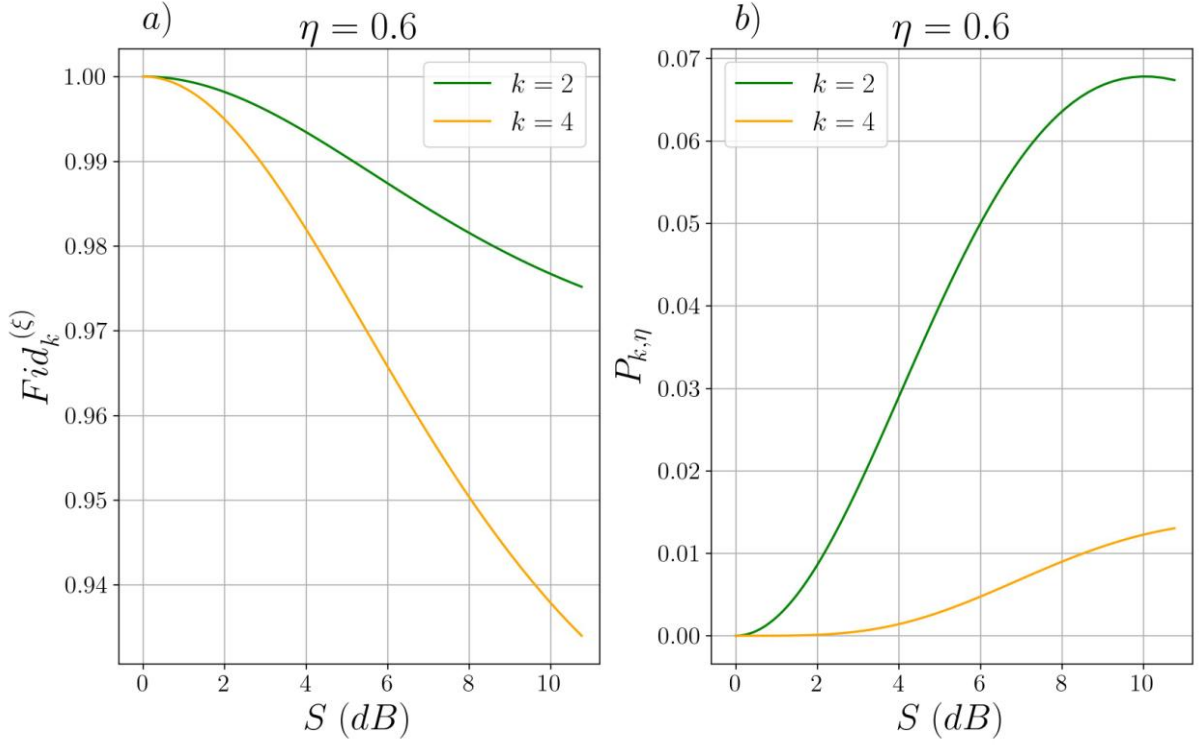

**Fig. 5(a,b).** (a) Dependence of the fidelity of the generated hybrid entanglement conditionally realized by the PNR detector with quantum efficiency $\eta = 0.6$ on the squeezing $S$ of the initial SMSV state for the case when an even number of photons are measured in the measuring mode. (b) Probabilities of the measurement outcomes $P_{k,\eta}$ in the case when PNR detector with $\eta = 0.6$ is used. The obtained probabilities take smaller values compared to those in Fig. 3(a) that could be obtained in the case of ideal PNR detector.